\documentclass[aps,apl,twocolumn,superscriptaddress]{revtex4-1}
\usepackage{amsfonts}
\usepackage{amsmath}
\usepackage{amssymb}
\usepackage{graphicx}
\usepackage{xcolor}
\usepackage{footmisc}

\begin{document}

\title{Sub-nanosecond spin-torque switching of perpendicular magnetic tunnel junction nanopillars at cryogenic temperatures}

\author{L. Rehm}
\email{laura.rehm@nyu.edu}
\affiliation{Center for Quantum Phenomena, Department of Physics, New York University, New York, NY 10003, USA}
\author{G. Wolf}
\affiliation{Spin Memory Inc., Fremont, CA 94538, USA}
\author{B. Kardasz}
\affiliation{Spin Memory Inc., Fremont, CA 94538, USA}
\author{M. Pinarbasi}
\affiliation{Spin Memory Inc., Fremont, CA 94538, USA}
\author{A. D. Kent}
\email{andy.kent@nyu.edu}
\affiliation{Center for Quantum Phenomena, Department of Physics, New York University, New York, NY 10003, USA}

\date{\today}

\begin{abstract}
Spin-transfer magnetic random access memory is of significant interest for cryogenic applications where a persistent, fast, low-energy consumption and high device density is needed. Here we report the low-temperature nanosecond duration spin-transfer switching characteristics of perpendicular magnetic tunnel junction (pMTJ) nanopillar devices (40 to 60\,nm in diameter) and contrast them to their room temperature properties. Interestingly, at fixed pulse voltage overdrive the characteristic switching time decreases with temperature, in contrast to macrospin model predictions, with the largest reduction in switching time occurring between room temperature and 150\,K. The switching energy increases with decreasing temperature, but still compares very favorably to other types of spin-transfer devices at 4\,K, with $<$ 300\,fJ required per switch. Write error rate (WER) measurements show highly reliable (WER $\leq$ 5$\times$10\textsuperscript{-5} with 4\,ns pulses at 4\,K) demonstrating the promise of pMTJ devices for cryogenic applications and routes to further device optimization.
\end{abstract}
\maketitle
Spin-transfer torque (STT) magnetic memory elements are interesting for cryogenic applications, such as computing systems based on superconducting circuits~\cite{Holmes2013}, because they are fast, energy efficient, have a small footprint and offer non-volatile data storage~\cite{Slonczewski1996,Berger1996}. STT memory devices typically consist of two thin ferromagnetic layers, one with a magnetization free to reorient and the other with a fixed magnetization direction both with perpendicular anisotropy separated by a thin insulating barrier. The memory states are layer magnetizations aligned either parallel (P) or antiparallel (AP). Intense commercial interest has led to the optimization of perpendicular magnetic tunnel junction (pMTJ) devices and materials that function near and even above room temperature~\cite{KentWorledge2015}. However, pMTJ device characteristics have not been studied in detail at low temperature.

Recently, three-terminal cryogenic spin-Hall-based memory devices have been demonstrated~\cite{Nguyen2019}. While these devices were optimized for low temperature operation and integration with superconducting circuitry, a two-terminal pMTJ device has advantages in terms of the integration density and simplicity of fabrication. Different two-terminal STT all-metallic magnetic memory elements~\cite{Rowlands2019,Rehm2019} have also been investigated at low temperature. They have a lower impedance, but they do not simultaneously offer high switching probabilities and large readout signals, i.e. large magnetoresistance. In addition to the advantages already mentioned, pMTJs offer long-term data storage even for nanopillar junctions just 10 nanometers in diameter and large tunnel magnetoresistance (TMR)~\cite{Park2012,Jan2014,Ikeda2010}. 

In conventional STT-MRAM devices operating at or above room temperature, the angle between the magnetization of layers is always non-zero. This reduces the time required to reverse the magnetization and hence reduces write errors. In other words, elevated temperature helps the write process but at the same time reduces the data retention time. For cryogenic memory devices, on the other hand, a simple macrospin model predicts that in absence of temperature the switching time would increase substantially~\cite{Sun2000}. But this prediction has not been tested in state-of-the-art perpendicularly magnetized magnetic tunnel junctions.   

In this letter we report the low-temperature high-speed spin-transfer switching characteristics of pMTJs and compare them to those at room temperature. We find that at a fixed pulse voltage overdrive the characteristic switching time decreases with temperature, in contrast to macrospin model predictions. The largest reduction in switching time occurs between room temperature and 150\,K. Further, at low temperatures there is a factor two increase in the device magnetoresistance, providing a much large readout signal. Remarkably, the write energies (103\,fJ, AP$\rightarrow$P and 286\,fJ, P$\rightarrow$AP at 4\,K) are much lower than devices with a metallic write channel, and thus a lower impedance~\cite{Nguyen2019}. Results on nanopillars as small as 40\,nm in diameter are presented, including write error rate (WER) measurements showing highly reliable (WER $\leq$ 5$\times$10\textsuperscript{-5} with 4\,ns pulses at 4\,K) demonstrating the promise of state-of-the-art pMTJ devices for cryogenic applications.

\begin{figure}
\includegraphics[width=0.48\textwidth,keepaspectratio]{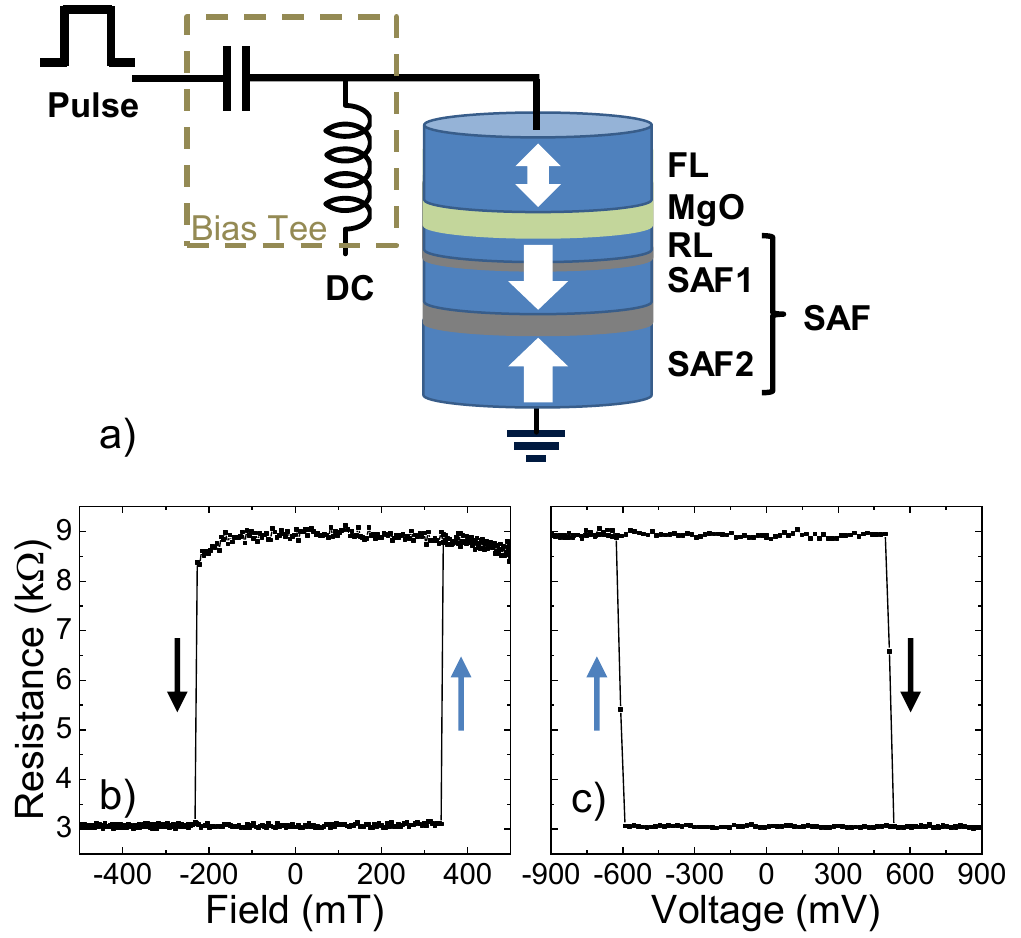}
\caption{a) Schematic of a pMTJ device and the pulse and readout measurement circuit. Nanosecond duration write pulses are applied through the capacitive port of a bias tee while the DC port is used for device read out. b) Resistance versus perpendicular field free layer hysteresis loop of a 40\,nm diameter device at 4\,K. The TMR ratio is 203\%. c) Voltage-induced switching with long duration (100\,ms) pulses of the same device at 4\,K in zero applied field. The junction resistance for the data in panels (b) and (c) is measured with a 30\,mV DC bias, a bias much less than the switching voltage.}
\label{Fig:Fig1}
\end{figure}

We studied pMTJ nanopillars with a perpendicularly magnetized CoFeB composite free layer (FL) consisting of CoFeB layer with a thin W insertion layer, CoFeB(1.5)/W(0.3)/CoFeB(0.8), where the numbers are the layer thicknesses in nm. The W insertion layer increases the perpendicular magnetic anisotropy and therefore enhances the thermal stability of the device~\cite{Sato2012,Kim2015}. This FL is one of the electrodes of a MgO tunnel junction. The other electrode is a CoFeB(0.9) reference layer (RL), which is ferromagnetically coupled to a first synthetic antiferromagnetic layer (SAF1) (see Fig.~\ref{Fig:Fig1}(a)). The synthetic antiferromagnetic layers (SAF) incorporate two antiferromagnetically coupled perpendicularly magnetized layers: (SAF1) [Pt(0.4)/Co(0.6)]$_{\times 2}$ and (SAF2) [Pt(0.4)/Co(0.6)]$_{\times 7}$; the full stack is SAF/RL(0.9)/MgO(1)/FL(2.6).
Following deposition, the wafer was annealed at 400$^\circ$C for 25 min. The annealed wafer was then pattered into circular-shaped nanopillars of 40, 50, and 60\,nm diameter using a combination of electron beam lithography and Ar ion beam milling. 

The devices are first characterized by measuring their field and current pulse resistance hysteresis loops. Figure~\ref{Fig:Fig1}(b) shows the minor hysteresis loops of a 40\,nm diameter pMTJ device measured in an applied perpendicular field at 4\,K. We observe sharp switching from P to AP states and vice versa with a field offset of  56\,mT, reflecting the fringe field from the SAF acting on the free layer~\cite{Gopman2012}. This sample exhibits a tunnel magnetoresistance (TMR) ratio of 203\% and an average coercive field of 283\,mT. Figure~\ref{Fig:Fig1}(c) shows voltage-induced switching of the same 40\,nm diameter device in zero field with 100\,ms duration voltage pulses. We observe a bistable region around zero applied voltage and voltage-induced switching with pulse amplitudes of 405\,mV for AP$\rightarrow$P switching and -358\,mV for P$\rightarrow$AP switching. Table I shows the TMR values extracted from the pulsed voltage loops from 4 to 295\,K. We observe almost a factor of two increase of the TMR at 4\,K compared to its value at room temperature, consistent with earlier studies~\cite{Lu1998,Cao2018}.

High speed spin-torque switching was studied by applying current pulses less than 5\,ns in duration using a pulse generator (Picosecond Pulse Labs 10,070A). A DAC board (National Instruments PCIe-6353) was used to apply longer (10\,$\mu$s) duration pulses to set and reset the magnetization direction of the free layer. 
The state of the device is again determined by applying a small voltage (30\,mV) with the DAQ board and measuring the resulting junction current. We use a bias-tee (Picosecond Pulse Labs 5575A) to combine low-frequency measurements with the DAQ with nanosecond pulses (see Fig.~\ref{Fig:Fig1}(a)). All measurements are performed in a cryogenic probe station where the sample stage can be heated up to 150\,K. Room temperature measurements are performed in the same setup with the cryostat cold head turned off. 

\begin{figure}
\includegraphics[width=0.48\textwidth,keepaspectratio]{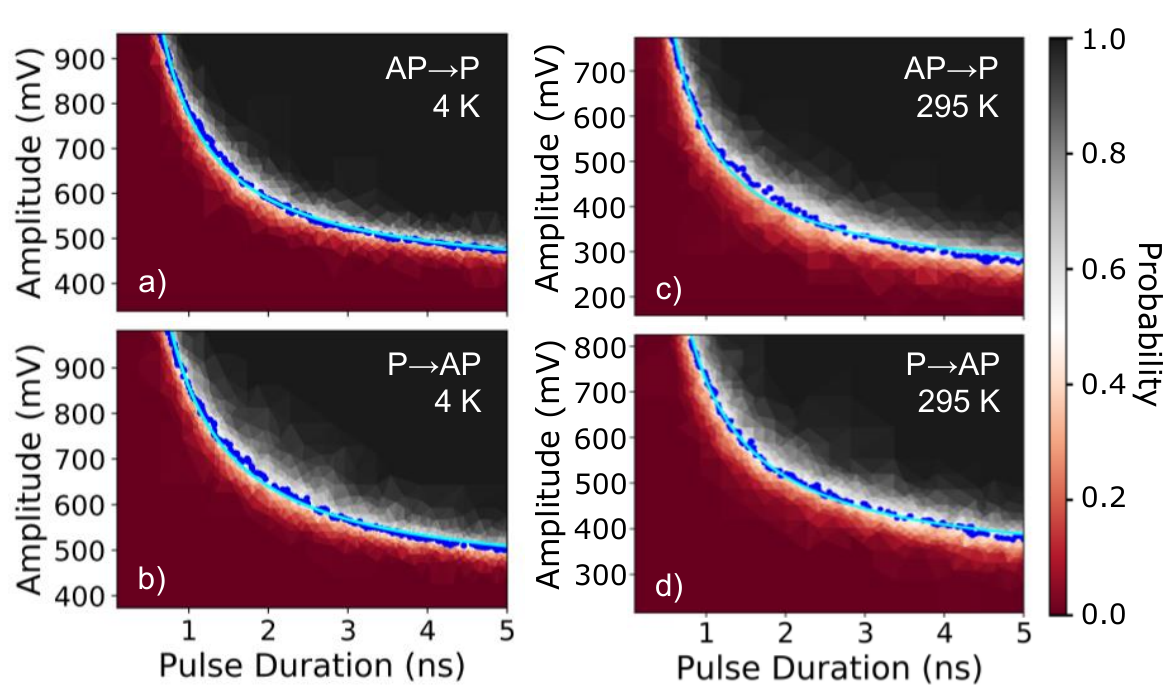}
\caption{Nanosecond pulsed current switching results at 4 and 295\,K.  Switching phase diagrams of a 40\,nm diameter pMTJ at 4\,K, a) AP$\rightarrow$P and b) P$\rightarrow$AP, and 295\,K, c) AP$\rightarrow$P and d) P$\rightarrow$AP. The color in the plot represents the switching probability, where red corresponds to 0\% and black is 100\%. The blue points represent the 50\% switching probability and the solid cyan line shows the fit to the macrospin model described in the main text.}
\label{Fig:Fig2}
\end{figure}

The measurement procedure thus consists of applying two square pulses---reset and write pulses---with opposite pulse polarities and reading the junction resistance and thus the junction state (P or AP) after each pulse. We start by applying a reset pulse to bring the device to a known state, either P or AP. We then verified the desired state by measuring the device resistance. The subsequent write pulse is applied by the pulse generator and the end state is determined by measuring the device resistance. The whole procedure is repeated about 100 times for each write pulse amplitude and duration combination to determine the switching probability. We systematically vary the pulse amplitude and duration to create the phase diagrams shown in Fig.~\ref{Fig:Fig2}. We focused our measurements on the most information rich area of the phase diagram, the vicinity of the 50\% switching probability boundary, by employing an adaptive measuring strategy~\cite{Nguyen2018}. We performed pulse measurements at 4, 75, 150, and 295\,K; the 4 and the 295\,K phase diagrams are shown in Fig.~\ref{Fig:Fig2}. 

\begin{figure}
\includegraphics[width=0.48\textwidth,keepaspectratio]{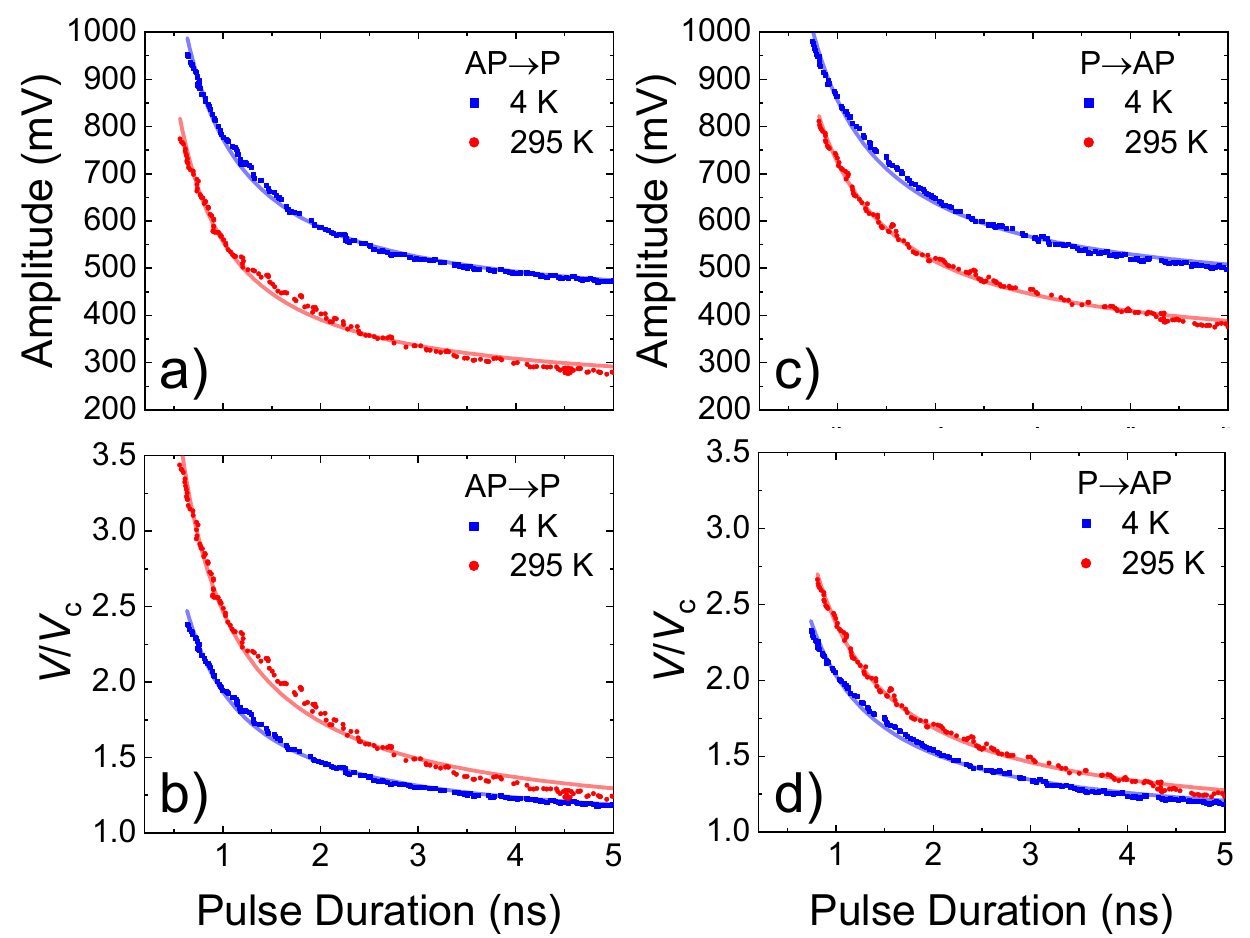}
\caption{50\% switching probability boundary of the same 40\,nm diameter device for a) AP$\rightarrow$P and c) P$\rightarrow$AP switching directions for 4 and 295\,K. b) and d) At fixed overdrive V/V\textsubscript{c} the device switches faster at 4\,K than at room temperature for both switching polarities. The lines show the fit to the macrospin model described in the main text.}
\label{Fig:Fig3}
\end{figure}

Figure~\ref{Fig:Fig2} shows the switching phase diagrams for AP$\rightarrow$P (left panels) and P$\rightarrow$AP transitions (right panels) for a 40\,nm diameter pMTJ at 4\,K (Figs.~\ref{Fig:Fig2}(a) and (b)) and 295\,K (Fig.~\ref{Fig:Fig2}(c) and (d)) in zero applied field. We observe high switching probability for pulse durations less than 1\,ns from room temperature to 4\,K. For $\sim$5\,ns pulse durations, switching of the pMTJ at 4\,K occurs for higher pulse amplitudes compared to that at room temperature as shown in Figs.~\ref{Fig:Fig3}(a) and (c), especially for the AP$\rightarrow$P transition (Fig.~\ref{Fig:Fig3}(a)).  

In order to characterize the data trends we consider a macrospin model, a simple model that provides analytic expressions for the switching times in the ballistic limit and how they vary with material and device parameters~\cite{Sun2000,Koch2004}. In this model the threshold voltage for spin-transfer switching is given by:
\begin{equation}
V=V_c\left(1+\frac{\tau_0}{\tau}\right),
\label{eq:1}
\end{equation}
where $\tau$ is the pulse duration, $\tau_0$ is the characteristic time for switching and $V_c$ is the switching voltage in the long pulse duration limit.
The fits of the switching boundary (i.e. the 50\% switching probability versus pulse duration) to Eq.~\ref{eq:1} are displayed as blue lines in Fig.~\ref{Fig:Fig2} and the corresponding fit parameters are listed in Table I. The fit parameters for 75 and 150\,K are also shown in Table I. 

In Figs.~\ref{Fig:Fig3}(b) and (d) we compare the data at 4 and 295\,K directly by plotting the normalized pulse voltages versus pulse duration. We thus see that at fixed pulse voltage overdrive, $V/V_c$, the switching time has decreased at 4\,K relative to that at 295\,K. The same behavior was observed in the 50 and 60\,nm diameter pMTJ nanopillars. 

The threshold voltage in the macrospin model is given by:
\begin{equation}
V_c= \frac{2 \alpha e \mathcal{A}R t \mu_0 M_s H_\mathrm{k,eff}}{P \hbar},
\label{eqvc}
\end{equation}
where $\mathcal{A}$ is the disk area, $R$ is a device resistance,  $t$ is the free layer thickness,  $P$ is the spin polarization and $H_\mathrm{k,eff}$ is the effective perpendicular magnetic anisotropy. $H_\mathrm{k,eff}=2K/(\mu_0 M_s)-M_s$, the perpendicular anisotropy associated with spin-orbit coupling $K$ minus the demagnetization field $M_s$.
\begin{table*}
\caption{TMR and fit parameters from the pulsed switching measurements for various temperatures and the corresponding optimal write energies.}
\centering
\label{t:Fit parameters}
\setlength{\tabcolsep}{10pt}
\begin{tabular}{lccccccc}
\noalign{\smallskip} \hline \hline \noalign{\smallskip}
T (K) & TMR (\%) &\multicolumn{2}{c}{\textit{V}\textsubscript{c} (mV)} & \multicolumn{2}{c}{$\tau$\textsubscript{0} (ns)} & \multicolumn{2}{c}{\textit{E} (fJ)} \\
 & & AP$\rightarrow$P & P$\rightarrow$AP & AP$\rightarrow$P & P$\rightarrow$AP & AP$\rightarrow$P & P$\rightarrow$AP\\
\hline
4 & 200 & 399 & 421 & 0.94 &1.03 & 103 & 286 \\
75  & 193 & 393 & 416 & 0.94 &1.05 & 98 & 283 \\
150 & 182 & 381 & 403 & 0.96 &1.10 & 94 & 287 \\
295  & 117 & 225 & 305 & 1.48 &1.38 & 51 & 195 \\
\noalign{\smallskip} \hline \noalign{\smallskip}
\end{tabular}
\end{table*}

In our fits to the data we find that $V_c$ increases with decreasing temperature and saturates at temperatures less than about 150\,K for both AP$\rightarrow$P and P$\rightarrow$AP transitions (see Table I). Eq.~\ref{eqvc} shows that $V_c$ depends on several material parameters that can vary with temperature, notably, $\alpha$, $M_s$, $H_\mathrm{k,eff}$ and $P$~\cite{Yu2012,Iwata2018,Sato2018,Mohammadi2019}.
Specifically, the magnetization and magnetic anisotropy both increase with decreasing temperature. The increase in TMR at low temperature also suggests that the spin polarization increases with decreasing temperature. The increase in the spin-polarisation therefore counteracts the increase in magnetization and magnetic anisotropy and this, at least qualitatively, can explain the saturation of $V_c$ below 150\,K. 

The characteristic switching time scale $\tau_0$ is given by
\begin{equation}
\tau_0=\frac{1+\alpha^2}{\alpha \gamma \mu_0 H_\mathrm{k,eff}} \ln(2/\theta_0),
\label{Eq:tau0}
\end{equation}
where $\alpha$ is the damping, $\gamma$ is the gyromagnetic ratio, $\mu_0$ is the permittivity of free space, and $\theta_0$ is the average magnetization initial angle. $\theta_0$ is related to the temperature and the energy barrier to magnetization switching by $\theta_0 \approx 1/(2 \sqrt{\pi \Delta})$, where $\Delta$ is the ratio of the energy barrier to reversal to the thermal energy, $\Delta=E_b/(k_BT)$. Hence $\tau_0$ is, again, related to material parameters that depend on temperature:
\begin{equation}
\tau_0=\frac{1+\alpha^2}{\alpha \gamma \mu_0 H_\mathrm{k,eff}} \ln(4\sqrt{\pi E_b/(k_BT)}).
\label{Eq:tau0T}
\end{equation}
As $V_c$ is independent of temperature below about 150\,K it seems reasonable to assume that the relevant junction material parameters, like $E_b$ and $\alpha$, are also independent of temperature below 150\,K. Eq.~\ref{Eq:tau0T} then predicts that the characteristic switching time would increase by about 20\% at 4\,K relative to its value at 150\,K. This predicted behavior reflects the decrease in magnetization fluctuations as the temperature decreases. This behavior is clearly not seen experimentally and shows that the macrospin model cannot explain the low temperature data trends. 

\begin{figure}
\includegraphics[width=0.48\textwidth,keepaspectratio]{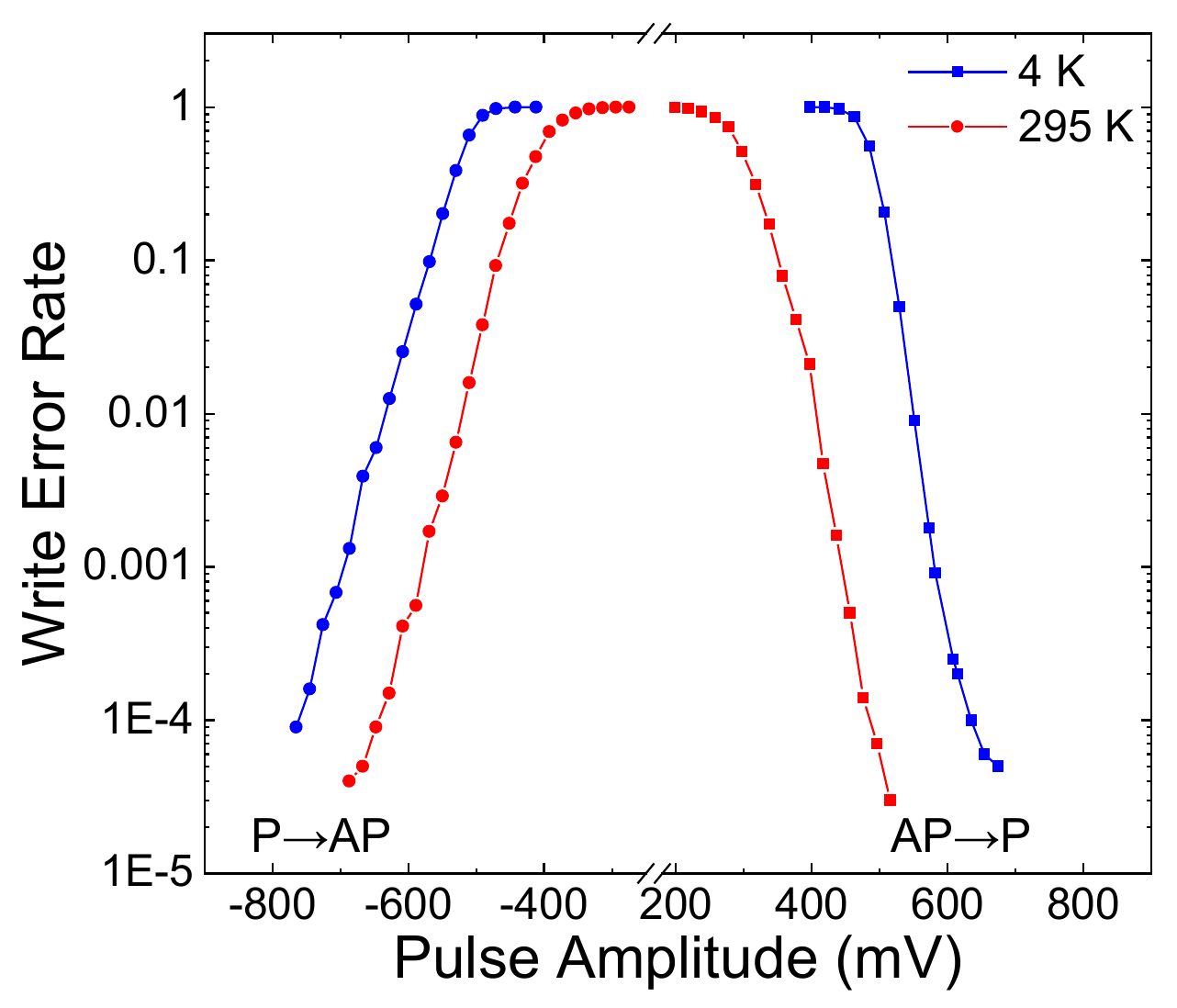}
\caption{Write error rates versus pulse amplitude for 4\,ns pulses at 4 and 295\,K of the identical 40\,nm diameter pMTJ device.}
\label{Fig:Fig4}
\end{figure}

In addition to characterizing the temperature dependence of the characteristic switching time and switching threshold we have measured write error rates for nanosecond current pulses. 
Figure~\ref{Fig:Fig4} shows the WER  for 4\,ns duration pulses at 4 and 295\,K for the same 40\,nm diameter pMTJ nanopillar. As already seen in the phase diagrams, the switching voltages are larger at 4\,K than at room temperature. Limited only be the measurement time, we found WER as low as 5$\times$10\textsuperscript{-5} at 4\,K (AP$\rightarrow$P) and no noticeable change in the slope of the WER curves versus pulse amplitude between 4 and 295\,K. This is an important result that highlights the fact that WER characteristics are not significantly dependent on temperature; that temperature simply rescales the pulse amplitude needed to achieve a desired WER performance.

Bases on these results we can determine the write energies and compare them to other types of cryogenic magnetic memory devices. The optimal (i.e. lowest) write energy is for pulse durations at the characteristics time $\tau_0$~\cite{Liu2014} and given by $E = V^2\tau_0/R$ with $V=2V_c$ and $R$ the device resistance at the switching voltage~\cite{Lu1998,Cao2018}. 
We find 103\,fJ for AP$\rightarrow$P and 286\,fJ for P$\rightarrow$AP at 4\,K for 40\,nm diameter devices. For larger diameter devices the optimal switching energy increases: for 50\,nm diameter devices, AP$\rightarrow$P 167 and P$\rightarrow$ AP 451\,fJ and for 60\,nm diameter devices AP$\rightarrow$P 226 and P$\rightarrow$AP 610\,fJ. The increased write energies for the larger devices is mostly associated with the increase in the threshold current for switching.  As expected, we observe a reduction of the optimal write energies with increasing temperature; that is,  thermal energy reduces the device switching energy (see Table I). 

These results clearly show an advantageous scaling of the switching energy with device diameter, the switching energy decreases as the device size decreases. These write energies are also comparable to write energies of all metal spin-valves with a Permalloy free layer~\cite{Rehm2019} and orthogonal spin-transfer spin valve devices~\cite{Rowlands2019}, which have much lower resistances but significantly larger switching currents. Further, two terminal pMTJ switching energies are lower than spin-Hall-based devices~\cite{Nguyen2019}, because of their lower switching currents.

The low energy consumption of pMTJ devices at 4\,K as well as their increased switching speed makes them very interesting as a low-energy cryogenic data storage solution. Foremost, the increased device magnetoresistance at low temperatures makes it possible to further reduce the resistance area product of the magnetic tunnel junction (reducing the device resistance and therefore the write energy) while maintaining a fast readout. It is also possible to  significantly reduce the switching energy. The most straightforward way would be by further reducing  the perpendicular magnetic anisotropy (see Eq.~\ref{eqvc}), as the FL in these studies are thermally stable at room temperature ($\Delta> $ 26, determined by using a read disturb rate method described in Ref.~\cite{Heindl2011}), meaning their magnetic anisotropy can be further reduced while still maintaining stable magnetic states at 4\,K. The FL magnetic moment can also be reduced to decrease the switching current. In addition, lower damping FL materials are also desirable for this application. In summary, two-terminal pMTJs are very promising for cryogenic applications and there are straightforward paths to further device optimization.

\noindent
{\bf Note added:} We are aware of related research by Dr. Li Ye at the Shanghai Institute of Microsystem and Information Technology, Chinese Academy of Sciences. Their work focuses on the low-temperature switching characteristics of pMTJs at longer time scales and the relation between the magnetic anisotropy, the magnetization and the switching voltages at low temperatures. Their paper will also be posted on the arXivs.

\begin{acknowledgments}
This research is supported in part by Spin Memory Inc. It is also based on work partially supported by the Office of the Director of National Intelligence (ODNI), Intelligence Advanced Research Projects Activity (IARPA), via contract W911NF-14-C0089. The views and conclusions contained herein are those of the authors and should not be interpreted as necessarily representing the official policies or endorsements, either expressed or implied, of the ODNI, IARPA, or the U.S. Government. The U.S. Government is authorized to reproduce and distribute reprints for Governmental purposes notwithstanding any copyright annotation thereon. This document does not contain technology or technical data controlled under either the U.S. International Traffic in Arms Regulations or the U.S. Export Administration Regulations. 
\end{acknowledgments}

%

\end{document}